\documentclass[aps,amsmath,amssymb]{revtex4}
\usepackage{graphicx}

\def\eq#1{Eq.~(\ref{#1})}
\def\fig#1{Fig.~\ref{#1}}

\begin{document}

\title{Adhesion clusters under shared linear loading: a stochastic analysis}
\author{T. Erdmann}
\affiliation{Max Planck Institute of Colloids and Interfaces, 14424 Potsdam, Germany}
\author{U. S. Schwarz}
\affiliation{Max Planck Institute of Colloids and Interfaces, 14424 Potsdam, Germany}

\begin{abstract}
We study the cooperative rupture of multiple adhesion bonds under
shared linear loading. Simulations of the appropriate Master equation
are compared with numerical integration of a rate equation for the
mean number of bonds and its scaling analysis. In general,
force-accelerated rupture is rather abrupt. For small clusters and
slow loading, large fluctuations occur regarding the timepoint of
final rupture, but not the typical shape of the rupture
trajectory. For vanishing rebinding, our numerical results confirm
three scaling regimes predicted before for cluster lifetime as a
function of loading rate. For finite rebinding, the intermediate
loading regime becomes irrelevant, and a sequence of two new scaling
laws can be identified in the slow loading regime.
\end{abstract}

\maketitle

\section{Introduction}

Cell adhesion is based on a large variety of different adhesion
molecules, each of which is optimised for its specific biological
function \cite{c:bell78}.  Most adhesion molecules have evolved to
operate under force. For example, in cell-matrix adhesion, receptors
from the integrin family usually function under conditions of cellular
contractility \cite{c:geig02}, while in leukocyte adhesion to blood
vessel walls, receptors from the selectin family operate under shear
flow \cite{c:ross97}. During recent years, single molecule force
spectroscopy has revealed formerly hidden properties of many different
adhesion molecules, which in the future might be linked explicitly to
their biological function \cite{c:evan01a}. Rupture of molecular bonds
under force is a stochastic process which can be modeled with Kramers
theory as thermally activated escape over a sequence of transition
state barriers \cite{c:evan97,c:izra97,c:seif98}.  The most convenient
loading protocol is a linear ramp of force, both experimentally and
theoretically.  For single molecules, the most frequent rupture force
as a function of the logarithm of loading rate has been predicted to
be a sequence of linear parts, each of which corresponding to one
transition state barrier along the rupture path \cite{c:evan97}. This
prediction has been confirmed experimentally for many different
adhesion systems \cite{c:merk99,c:evan01a}, including $\alpha_5
\beta_1$-integrin \cite{c:li03} and L-selectin mediated bonds
\cite{c:evan01b}.

Although single molecule force spectroscopy has strongly changed our
understanding of specific adhesion, cell adhesion is usually not based
on single molecules, but on clusters of varying size. Therefore, future
understanding of cell adhesion also has to include the cooperative
behaviour of adhesion molecules under force. In single molecule
experiments, ruptured bonds usually cannot rebind due to elastic
recoil of the transducer. In contrast, ruptured bonds in adhesion
clusters can rebind as long as other bonds are still closed, thus
holding ligands and receptors in close proximity. For adhesion
clusters under constant loading, it is well known that despite
rebinding, stability is lost beyond a critical force
\cite{c:bell78}. For adhesion clusters under linear loading, force
grows without bounds and the cluster will always rupture. Recently,
the most frequent rupture force has been measured as a function of
loading rate for clusters of $\alpha_{\nu} \beta_3$-integrins and
RGD-lipopeptides loaded through a soft transducer in a homogeneous way
\cite{c:prec02}. Theoretically, it has been shown before that
different scaling regimes exist for cluster lifetime as a function of
cluster size, loading rate and rebinding rate
\cite{c:seif00}. For the case of a stiff transducer, force on single
bonds is independent of the number of closed bonds and a mean field
approximation can be applied to make further theoretical progress
\cite{c:seif02}. However, for the case of a soft transducer, force is
shared between closed bonds, leading to real cooperativity: if one of
the closed bonds ruptures, force is redistributed over the remaining
closed ones. Here we present for the first time a full treatment of
this case. We start with a one-step Master equation with Kramers-like
rates, which is solved by Monte Carlo methods. These results are then
compared to numerical integration of a rate equation for the mean
number of bonds. We show that considerable differences exist between
the stochastic and deterministic treatments for small clusters or slow
loading. For the case of vanishing rebinding, our results confirm the
three scaling regimes for cluster lifetime as a function of loading
rate, which have been predicted before on the basis of a scaling
analysis of the rate equation for the mean number of bonds
\cite{c:seif00}. For the case of finite rebinding, the intermediate
scaling regime becomes irrelevant. For slow loading, we identify a
sequence of two new scaling laws, which result from stochastic decay
towards an absorbing boundary and finite rupture strength at constant
loading, respectively.

\section{Model}

We consider a cluster with $N_t$ parallel bonds. At any time $t$, $i$
bonds are closed and $N_t-i$ bonds are open ($0 \le i \le N_t$). The
$i$ closed bonds are assumed to share force $F$ equally, that is each
closed bond is subject to the force $F/i$. In the following, we will
consider linear loading, that is $F = r t$ where $r$ is loading rate.
Single closed bonds are assumed to rupture with the dissociation rate
$k = k_0 e^{F/i F_b}$, which corresponds to the case of one sharp
transition state barrier along the rupture path
\cite{c:bell78,c:evan97}.  Here, $F_b$ is the internal force scale of
the bond set by the barrier. Single open bonds are assumed to rebind
with the force independent association rate $k_{on}$.  We now
introduce dimensionless variables: dimensionless time $\tau = k_0 t$,
dimensionless loading rate $\mu = r / k_0 F_b$ and dimensionless
rebinding rate $\gamma = k_{on}/k_0$. The stochastic dynamics of our
model is described by a one-step Master equation
\begin{equation} \label{MasterEquation}
\frac{dp_i}{d\tau} = r_{i+1} p_{i+1} + g_{i-1} p_{i-1} - [ r_i + g_i ] p_i
\end{equation}
where $p_i(\tau)$ is the probability that $i$ closed bonds are present at
time $\tau$. The reverse and forward rates between the different states 
$i$ follow from the single molecule rates as
\begin{equation} \label{Rates}
r_i = i e^{\mu \tau / i}\quad\text{and}\quad g_i = \gamma (N_t - i)\ .
\end{equation}
For constant force, this Master equation has been studied before
\cite{c:coze90,uss:erdm04a}. Since adhesion clusters (like single
molecules) usually cannot rebind from the completely dissociated state
due to elastic recoil of the transducer, we implement an absorbing
boundary at $i = 0$ by setting $g_0 = 0$. Since force increases in
time without bounds, the cluster will always dissociate in the long
run, that is $p_i(\tau) \to \delta_{i0}$ for $\tau \to \infty$, both
for absorbing and reflecting boundaries. Cluster lifetime $T$ is the
mean time to reach the absorbing state $i = 0$.  By defining cluster
dissociation rate $D = dp_0/d\tau = r_1 p_1$, cluster lifetime follows
as $T = \int_0^{\infty} d\tau\ \tau D$.  Since the reverse rates $r_i$
are non-linear in $i$ and time-dependent, an analytical solution
for the $p_i$ as a function of the three model parameters $N_t$, $\mu$
and $\gamma$ seems to be impossible. Therefore we solve the Master
equation numerically using the Gillespie algorithm for efficient Monte
Carlo simulations, typically averaging over $10^5$ simulation
trajectories for each set of parameters \cite{c:gill77}.

A quantity of large interest is the mean number of closed bonds, $N =
\langle i \rangle = \sum_{i=1}^{N_t} i p_i$. In a continuum approach,
one expects that this quantity satisfies the ordinary differential
equation
\begin{equation} \label{DeterministicEquation}
\frac{dN}{d\tau} = - N e^{\mu \tau / N} + \gamma (N_t - N)\ .
\end{equation}
Cluster lifetime $T$ can be defined by $N(T) = 1$. Several different
scaling regimes for $T$ as a function of $N_t$, $\mu$ and $\gamma$
have been predicted on the basis of \eq{DeterministicEquation}
\cite{c:seif00}. Below these scaling predictions will be compared to
both numerical integration of the deterministic equation and to our
stochastic results.

\section{Decay without rebinding}

\begin{figure}
\includegraphics{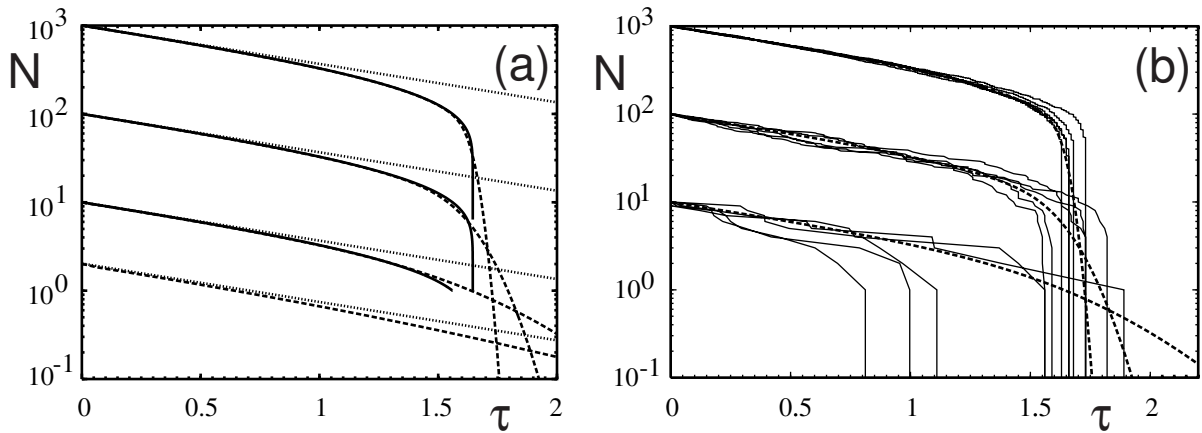}
\caption{(a) Mean number of closed bonds $N$ as a function of time $\tau$ 
for the case of vanishing rebinding, $\gamma = 0$, for $\mu / N_0 =
0.1$ and $N_t = N_0 = 2, 10, 10^2$ and $10^3$. Dotted lines: Initial
exponential decay. Dashed lines: first moment of the Monte Carlo
simulations.  Solid lines: numerical integration of deterministic rate
equation (not for $N_t = 2$). (b) Individual trajectories from Monte
Carlo simulations in comparison with the mean (not for $N_t = 2$).}
\label{fig:N_ga=0}
\end{figure}

We first consider the case of vanishing rebinding, $\gamma = 0$.
In this case the total number of bonds $N_t$ does not appear in the
model equations and the initial condition $N(0) = N_0$ is the only
relevant parameter concerning the number of bonds. The scaling
analysis of \eq{DeterministicEquation} suggests that decay can be
divided into two parts \cite{c:seif00}. Initial decay is not yet
affected by loading and thus is exponential with $N(\tau) = N_0
e^{-\tau}$. The second part of the decay is super-exponential and can
be shown to be much shorter than the first one. Therefore the
crossover time, which is defined by an implicit function, determines
cluster lifetime $T$. In the regime of slow loading, $\mu < 1$,
exponential decay persists until $N(\tau) = 1$ and $T = \ln N_0$. In
the regime of intermediate loading, $1 < \mu < N_0$, the crossover
occurs before $N(\tau) = 1$ is reached, and lifetime is reduced to $T
\sim \ln (N_0 / \mu)$. In the regime of fast loading, $\mu > N_0$,
lifetime scales even stronger with loading rate, $T \sim (N_0 / \mu)
\ln (\mu / N_0)$.

In \fig{fig:N_ga=0}a we plot $N(\tau)$ as obtained from simulations of
the Master equation (dashed lines) and from numerical integration of
the deterministic equation (solid lines) for $N_0 = 2, 10, 10^2$ and
$10^3$. The dotted lines are the exponential decays $N(\tau) = N_0
e^{-\tau}$ for vanishing loading. In the presence of loading, the
later part of the decay process clearly is super-exponential.  The
first moment of the stochastic process decays less abrupt than the
deterministic result, although for increasing cluster size, the
difference between stochastic and deterministic results becomes
smaller. In \fig{fig:N_ga=0}b, we show representative trajectories
from Monte Carlo simulations. They demonstrate that the final stage of
the rupture process is rather abrupt. In fact abrupt decay is typical
for shared loading and is found also for shared \emph{constant}
loading: a decreasing number of closed bonds increases force on the
remaining bonds, thus further increasing their dissociation rates
\cite{uss:erdm04a}. As \fig{fig:N_ga=0}b shows, fluctuations tend to
change the timepoint of rupture, rather than the typical shape of the
decay curve.  For increasing cluster size, fluctuations become smaller
and rupture events are concentrated around the rupture of the
deterministic cluster. An analysis of the variance of the number of
closed bonds $i$ shows that for slow loading, it is close to the exact
result for vanishing loading, $\langle i^2 \rangle - \langle i
\rangle^2 = N_0 e^{- \tau} (1 - e^{-\tau})$ \cite{r:mcqu63}. It
vanishes for $\tau = 0$ due to the initial condition, then quickly
rises to a maximum and finally decays exponentially. As loading rate
$\mu$ increases, a large additional peak appears shortly before final
rupture (not shown).

\begin{figure}
\includegraphics{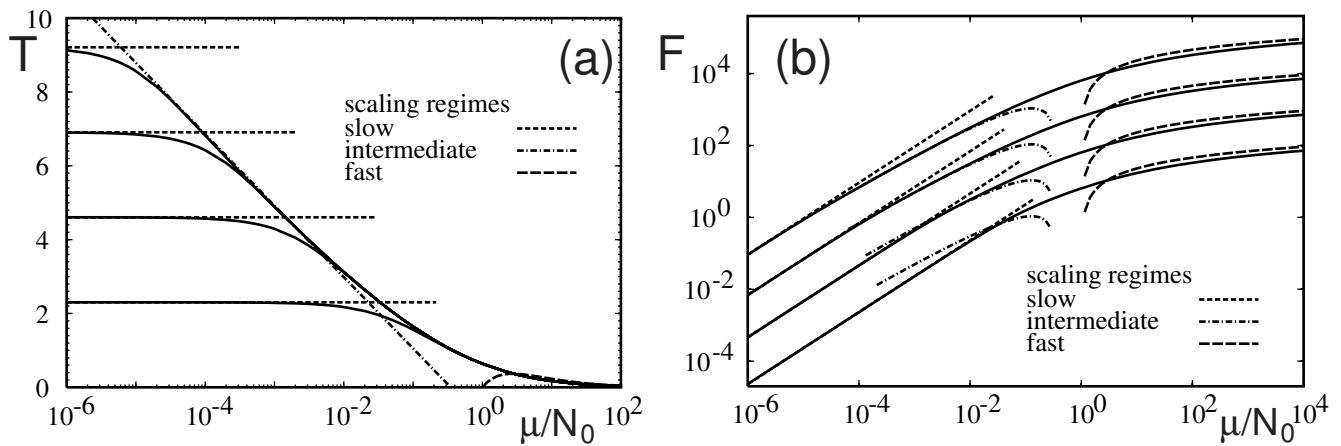}
\caption{Solid lines: deterministic results for (a) cluster lifetime $T$ and 
(b) rupture force $F = \mu T$ for the case of vanishing
rebinding, $\gamma = 0$, as a function of $\mu / N_0$ for $N_t = N_0 = 10,
10^2, 10^3$ and $10^4$. Dashed lines: curves for all three scaling
regimes.}
\label{fig:T_ga=0}
\end{figure}

Simulations allow to measure cluster dissociation rate $D(\tau)$
and cluster lifetime $T$ for all parameter values. 
For $\mu < 1$, the simulation results are close to the known
analytical results for $\mu = 0$, $D(\tau) = N_0 e^{-\tau} (1 - e^{-\tau})^{N_0 - 1}$ 
\cite{r:mcqu63} and $T = \sum_{i=1}^{N_0} 1/i \approx \ln N_0 + 
(1/2 N_0) + 0.577$ \cite{c:gold96,c:tees01}. For large $N_0$, the
deterministic scaling $T = \ln N_0$ results. For $\mu > 1$, the
functions $D(\tau)$ become narrowly peaked around the mean value
$T$. As suggested by the scaling analysis, we find that now $T$
depends only on the value of $\mu / N_0$. In \fig{fig:T_ga=0}a, we
plot deterministic results for $T$ as a function of $\mu / N_0$ and
for different values of $N_0$. The stochastic results are very
similar, except for the differences in the initial plateau
values. Initially, the different curves plateau at the values $\ln
N_0$ for $\mu < 1$. For $1 < \mu < N_0$ and sufficiently large $N_0$,
they collapse onto a universal curve, which can be approximated by
$0.84 \ln (0.35 N_0/\mu)$. For $\mu > N_0$, they collapse onto another
universal curve, $(N_0/\mu) \ln (\mu / N_0)$. In \fig{fig:T_ga=0}b, we
plot the logarithm of the deterministic rupture force, $F = \mu T$, as a
function of $\mu / N_0$. For large $N_0$, one clearly sees the
sequence of the three different scaling regimes.  For decreasing
$N_0$, the intermediate scaling curve becomes an increasingly bad fit.

\section{Effect of rebinding}

\begin{figure}
\includegraphics{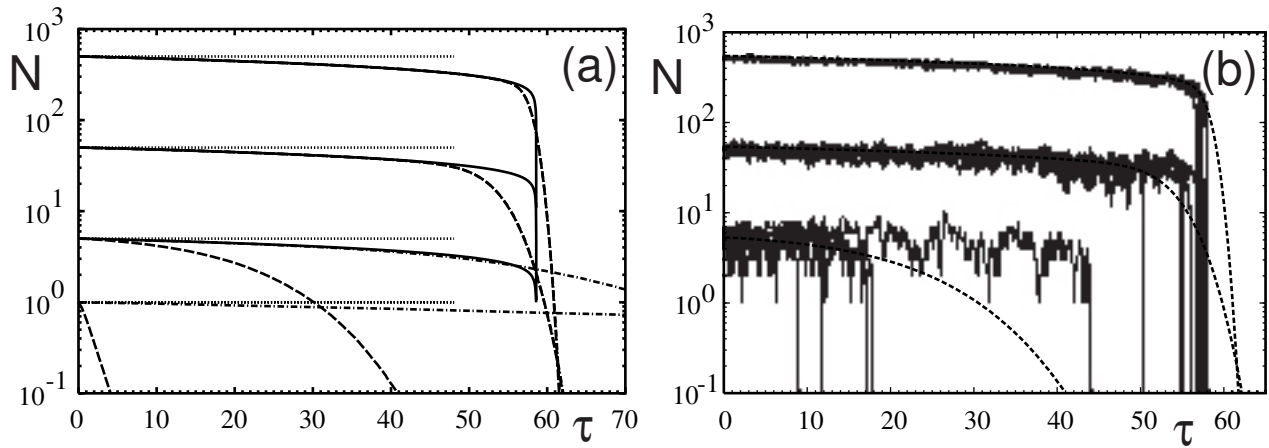}
\caption{(a) Mean number of closed bonds $N$ as a function of time $\tau$ for
rebinding rate $\gamma = 1$ and loading rate $\mu/N_0 =
0.01$. Cluster sizes $N_t = 2$, $10$, $10^2$ and $10^3$, initial
condition $N_0 = N_{eq} = \gamma N_t/(1+\gamma)$. Dotted lines:
Initial number of closed bonds. Dashed lines: first moment of the
Monte Carlo simulations. Solid lines: numerical integration of
deterministic equation (not for $N_t = 2$). Dashed-dotted lines:
effect of reflecting boundary for $N_t = 2$ and $10$. (b) Individual
trajectories from Monte Carlo simulations in comparison with the
mean (not for $N_t = 2$).}
\label{fig:N_ga=1}
\end{figure}

In the stochastic framework, cluster lifetime can be identified with
the finite mean first passage time of reaching the absorbing boundary
at $i = 0$. For $\mu = 0$ and $N_0 = N_t$, an exact result can be
obtained with the help of Laplace transforms \cite{uss:erdm04a}:
\begin{equation} \label{T_stoch}
T_{stoch} = \frac{1}{1+\gamma} \left( H_{N_t} 
+ \sum_{i=1}^{N_t} \binom{N_t}{i} \frac{\gamma^i}{i} \right)\ ,  
\end{equation}
where $H_{N_t} = \sum_{i=1}^{N_t}(1/i)$ is the {$N_t$}th harmonic
number. In the deterministic framework of \eq{DeterministicEquation}
and $\mu = 0$, an adhesion cluster with a total of $N_t$ molecular
bonds will equilibrate from any initial number of closed bonds $N_0$
to a stable steady state with $N_{eq}=\gamma N_t/(1+\gamma)$ closed
bonds. For convenience, in the following we will use $N_0 = N_{eq}$.
Then similar results follow for $T_{stoch}$ as given in \eq{T_stoch}.
A stability analysis of the deterministic equation
\eq{DeterministicEquation} for loading with a \textit{constant}
force $f = F/F_b$ shows that the steady state cluster size decreases
until stability is lost beyond a critical force $f_c = N_t \rm plog
(\gamma/e)$ \cite{c:bell78,uss:erdm04a}. Here the product logarithm
$\rm plog(a)$ is defined as the solution $x$ of $xe^{x} = a$. For
$\gamma < 1$, the critical force can be approximated as $f_c \approx
N_t \gamma/e$: it vanishes with $\gamma$ since without rebinding the
cluster decays by itself. For $\gamma > 1$, it can be approximated as
$f_c \approx 0.5 N_t \ln \gamma$, that is the critical force now is
only a weak function of rebinding. For slow loading, $\mu < 1$, the
adhesion cluster will follow the quasi-steady state until the critical
force $f_c$ is reached at the time $\tau_c = f_c / \mu$.  The
remaining time to rupture is smaller and thus the lifetime of the
adhesion cluster is close to
\begin{equation}
T_{det} = \frac{N_t}{\mu} \rm plog \frac{\gamma}{e}\ .
\end{equation}
It diverges with the inverse of loading rate in the limit of vanishing
$\mu$, as it is required by the existence of a stable steady state and
a finite rupture force $F = \mu T = f_c$. For intermediate loading, $1
< \mu < N_0$, a power-law behaviour $T \sim (N_0/\mu)^{1/2}$ has
been erroneously predicted in ref.~\cite{c:seif00}, as reported in
ref.~\cite{c:seif02}. For fast loading, $\mu > N_0$, rebinding can be
neglected and $T \sim (N_0/\mu) \ln (\mu/N_0)$ as in the previous
section.

In \fig{fig:N_ga=1}a $N(\tau)$ is plotted for $\gamma = 1$ as obtained
from Monte Carlo simulations (dashed lines) and from numerical
integration of the deterministic equation (solid lines). The
different initial conditions $N_0$ for $N_t = 2, 10, 10^2$ and $10^3$
are represented by the dotted lines. In \fig{fig:N_ga=1}b individual
trajectories from the simulations are compared to the stochastic
averages from \fig{fig:N_ga=1}a. For the small clusters, $N_t = 2$
and $10$, loading rate is so small that $f_c / \mu > T_{stoch}$. Then
$T \approx T_{stoch}$ and the clusters decay by themselves due to
stochastic fluctuations to the absorbing boundary (\textit{ultra-slow
regime}). The dash-dotted lines in \fig{fig:N_ga=1}a show the effect
of a reflecting boundary, which is rather dramatic for these small
cluster sizes. For the large clusters, $N_t = 10^2$ and $10^3$,
fluctuations are less probable until the force is close to
$f_c$. Therefore the individual clusters fluctuate around the
quasi-steady state and dissociate only close to the deterministic
cluster lifetime $T_{det}$. Due to the large force on a single bond at
$f_c$, the boundary has little influence here.  A detailed
analysis of the variance of $i$ confirms this description (not shown): for
the smallest cluster, when fluctuations dominate during the whole time
evolution, the variance shows a broad peak. For the larger clusters,
it develops a narrow peak around the mean rupture time.

\begin{figure}
\includegraphics{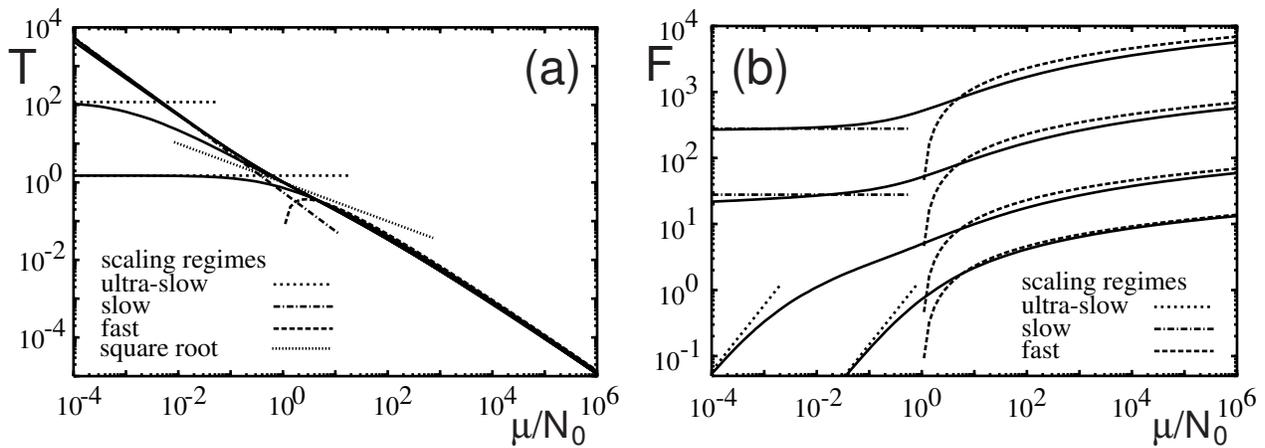}
\caption{(a) Mean cluster lifetime $T$ and (b) mean rupture force $F = \mu T$
for the case $\gamma = 1$ as a function of $\mu / N_0$ for $N_t
= 2, 10, 10^2$ and $10^3$. In (a), the curves for the two larger clusters
are nearly identical.}
\label{fig:T_ga_N}
\end{figure}

In \fig{fig:T_ga_N}, we show mean cluster lifetime $T$ and mean rupture
force $F = \mu T$ for $\gamma = 1$ as a function of $\mu / N_0$ for
the cases $N_t = 2, 10, 10^2$ and $10^3$. For the small clusters,
$T$ starts at the value of $T_{stoch}$ and ends in
the scaling regime for fast loading, where the curves are practically
identical for all different parameter values at a given value for
$\mu/N_0$. The curves for the large clusters are nearly
identical. They start at the values of $T_{det}$ for small loading
rates and end in the same fast loading regime. An intermediate loading
regime seems to exist only as a transient between the regimes of slow
and fast loading. In particular, it does not fit well to an inverse
square root dependence, as shown in \fig{fig:T_ga_N}a.

\section{Conclusions} In this paper, we have presented for the first
time a full analysis of the cooperative decay of a cluster of adhesion
bonds under linearly rising force.  Significant differences between
stochastic and deterministic treatments are found for small clusters
or slow loading, when stochastic fluctuations are relevant. However,
they do not affect so much the typical shape of the rupture
trajectory, but rather the timepoint at which rupture occurs.
For the case of vanishing rebinding, $\gamma = 0$,
our full treatment nicely confirms the scaling analysis of the
deterministic equation for cluster lifetime $T$ as a function of $\mu$
and $N_0$ \cite{c:seif00}. However, in contrast to the scaling analysis, the
full treatment presented here allows for detailed comparision with
experiments, e.g.\ in regard to typical unbinding trajectories or
binding strength over a range of loading rates spanning different
scaling regimes. For the case with finite rebinding, $\gamma > 0$, we
identify a sequence of two new scaling laws within the regime of slow
loading, $\mu < 1$. For ultra-slow loading, $T$ is independent of
$\mu$ and is determined by stochastic fluctuations towards the
absorbing boundary.  For larger $\mu$ (but still with $\mu < 1$), $T$
starts to scale inversely with $\mu$, due to the finite rupture
strength at constant loading. In contrast to the case of vanishing
rebinding, a scaling regime of intermediate loading, $1 < \mu <
N_0$, could not be identified. 

Our results can be applied for example to rolling adhesion of
leukocytes, when multiple L-selectin bonds are dynamically loaded in
shear flow \cite{uss:dwir03a}. Dynamic force spectroscopy has
only recently been applied to clusters of adhesion bonds
\cite{c:prec02}. RGD-lipopeptides on a vesicle have been presented to
$\alpha_{\nu} \beta_3$-integrins on a cell.  The effect of thermal
membrane fluctuations can be disregarded on both sides, because the
vesicle is under large tension and the integrins are rigidly
connected to the cytoskeleton. Appreciable loading occurs only over a
ring region along the rim of the contact disc, for which no
inhomogeneities have been observed. If one neglects the subsequent
peeling of the inner region, which presumably is much faster, our
model can be applied. The parameter values can be estimated to be $N_t
\approx 100$, $F_b \approx 40$ pN, $k_0 \approx 0.01$ Hz and $\gamma
\approx 1$.  Loading rates have been varied from $r = 20 - 4 \times
10^3$ pN/s, that is $\mu / N_t = 0.5 - 100$. Therefore this experiment
should correspond to the intermediate and fast loading regimes. We
expect that future improvements in experimentation will make it
possible to probe also the slow loading regime, where rebinding and
stochastic effects become relevant. In order to achieve a more
complete understanding of the role of force in cell adhesion, future
modeling should also address the detailed nature of the force
transducer, non-homogeneous loading and more realistic scenarios for
the rebinding process.

\acknowledgments
Helpful discussions with Udo Seifert and Rudolf Merkel are gratefully
acknowledged. This work was supported by the German Science Foundation
through the Emmy Noether Programme.

\end{document}